\documentclass{article}
\usepackage{spconfa4,amsmath,graphicx}
\usepackage{acronym}
\usepackage{siunitx}
\usepackage{caption}
\makeatletter
\def\section{\@startsection{section}{1}{\z@}%
  {1.4ex plus 0.6ex minus 0.2ex}%
  {0.8ex plus 0.2ex}%
  {\normalfont\normalsize\bfseries\centering}}
\def\subsection{\@startsection{subsection}{2}{\z@}%
  {1.1ex plus 0.5ex minus 0.2ex}%
  {0.6ex plus 0.2ex}%
  {\normalfont\normalsize\bfseries}}
\def\subsubsection{\@startsection{subsubsection}{3}{\z@}%
  {0.9ex plus 0.4ex minus 0.2ex}%
  {0.4ex plus 0.1ex}%
  {\normalfont\normalsize\itshape}}
\makeatother

\title{A Novel Binaural Cue Preservation Loss\\
for DNN-Based Binaural Speech Enhancement}
\name{Jayteerth Amble$^{1,2}$, Thomas Haubner$^{2}$, Hendrik Schr\"oter$^{2}$, Christoph Hoog Antink$^{1}$, Henning Puder$^{1,2}$}
\address{$^{1}$Technische Universit\"at Darmstadt, Darmstadt, Germany, \\$^{2}$ WSA, Erlangen, Germany}
\begin{document}
\acrodef{BSE}{binaural speech enhancement}
\acrodef{DNN}{deep neural network}
\acrodef{ILD}{interaural level differences}
\acrodef{IPD}{interaural phase differences}
\acrodef{HRIR}{head-related impulse response}
\acrodef{STFT}{short-time Fourier transform}
\acrodef{IBM}{ideal binary mask}
\acrodef{RTF}{relative transfer function}
\acrodef{BRE}{binaural reconstruction error}
\acrodef{MBSTOI}{modified short-time objective intelligibility}
\acrodef{DoA}{direction-of-arrival}

%
\maketitle
\begin{abstract}
Binaural speech enhancement for hearing aids aims to reduce noise while preserving the interaural cues needed for spatial localization. Although deep neural network-based methods achieve strong noise reduction, they often distort the relationship between the left and right signals. In this paper, we propose two novel binaural cue preservation losses. First, a binaural reconstruction error loss that directly penalizes masking-induced distortion in the relationship between the left and right spectra, providing a more direct measure of the binaural consistency than conventional separate \ac{ILD} and \ac{IPD} errors as in prior work. Second, a binaural cue loss that jointly models \ac{ILD} and \ac{IPD} to better preserve the binaural structure. Experimental results show that both proposed losses maintain strong noise reduction performance and reduce masking-induced distortion compared to the state-of-the-art baseline cue loss, while the second proposed joint binaural cue loss also outperforms the baseline in \ac{ILD} preservation.
\end{abstract}
%
\acresetall
\section{Introduction}
\label{sec:intro}

Noise reduction is a key component of modern speech enhancement systems for assistive listening devices such as hearing aids \cite{nr_ha1,nr_ha2} and AR/VR devices \cite{nr_ar}. Conventional single- and multi-channel algorithms estimate filters using statistical models of speech and noise \cite{beamforming,spectral_substraction}, but their performance often degrades in non-stationary environments.
More recently, \ac{DNN}-based methods have gained relevance due to their ability to generalize well to dynamically changing real-world scenarios. Many state-of-the-art single-channel methods operate either in the time domain \cite{td_1, td_2} or in the \ac{STFT} domain \cite{dccrn,Deepfilternet}. In the latter case, complex-valued time-frequency (T-F) masks are estimated to suppress noise. Although monaural approaches achieve strong denoising performance, applying them independently to the left and right channels can distort binaural cues and thereby degrade spatial perception.
In the \ac{STFT} domain, \ac{DNN}-based \ac{BSE} typically processes the complex spectra of the left and right microphone signals jointly to estimate a complex T-F mask for each channel \cite{bc_old,bc_transformer,bc_lstm}. In addition to noise suppression, binaural processing aims to preserve spatial cues, commonly characterized by \ac{ILD} and \ac{IPD}, which are essential for sound localization and speech understanding in noisy environments.
A key challenge is the trade-off between noise reduction and spatial cue preservation. Models optimized primarily for denoising can distort interaural cues. To address this issue, previous work introduced explicit binaural cue loss terms based on separate \ac{ILD} and \ac{IPD} distortion measures, steering the network towards preserving spatial information during training \cite{bc_transformer,bc_lstm,bc_fourier}. The addition of cue-preserving loss allows the network to use residual noise components to better preserve target spatial cues, but generally at the cost of reduced denoising performance. In addition, as there is no standardized way to quantify localization distortion directly, binaural distortion is commonly assessed using separate \ac{ILD} and \ac{IPD} error measures \cite{bc_transformer,bc_lstm,bc_fourier}. However, direct IPD comparisons are affected by phase wrapping, and together these binaural cue metrics do not fully capture masking-induced changes in the interchannel relationship. In this work, we make two contributions. First, we introduce a new \ac{BRE} loss that directly penalizes masking-induced distortion in the interchannel relationship. Second, we propose a new binaural cue preservation loss that jointly constrains \ac{ILD} and \ac{IPD}. The proposed formulation penalizes \ac{ILD} distortion while encoding \ac{IPD} on the complex unit circle, resulting in a joint binaural cue constraint that supports robust cue preservation while maintaining strong denoising performance.

\section{Signal model and proposed method}
In this section, we first describe the signal model. We then introduce the  baseline binaural cue loss used in state-of-the-art methods, followed by the proposed binaural cue losses and the noise reduction loss.

\subsection{Signal model}
We model the time-domain binaural microphone signal at sample index \(k\) and channel \(i \) as
\begin{figure*}[!t]
    \centering
    \includegraphics[width=\textwidth,height=0.25\textheight,keepaspectratio]{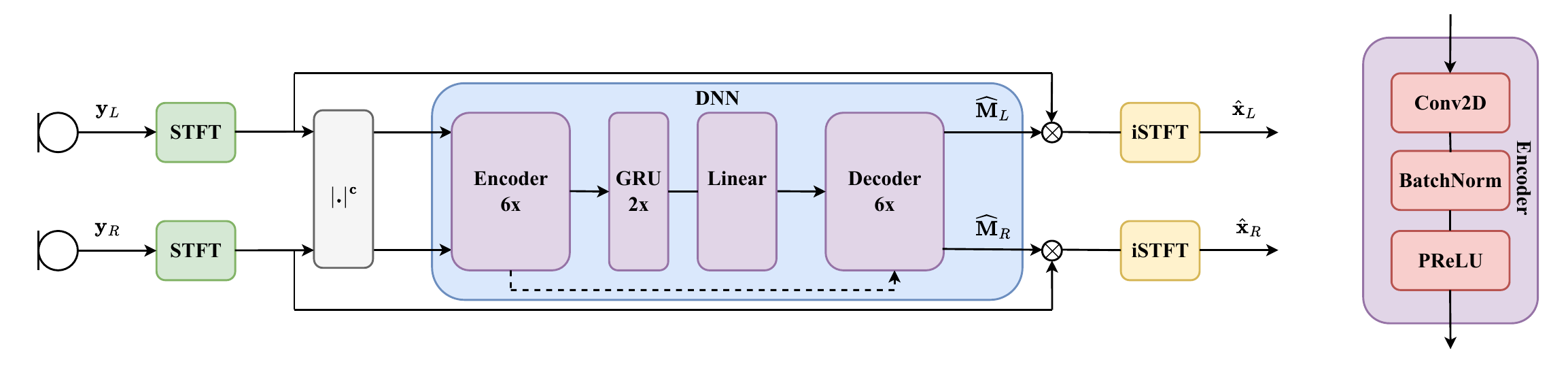}
    \caption{Block diagram of the proposed binaural speech enhancement framework. Skip connections are used between encoder and decoder. The decoder mirrors the encoder using transposed Conv2D layers.}
    \label{fig:block_diagram}
\end{figure*}
\label{sec:sig_model}
{\ninept\begin{equation}
\begin{aligned}
y_i(k) &= s(k) * h_i(k) + v_i(k) \\
       &= x_i(k) + v_i(k), \quad i \in \{\mathrm{L}, \mathrm{R}\} ,
\end{aligned}
\label{eq:time_domain_mixture}
\end{equation}}

where \(i\) denotes the left or right channel, \(x_i(k)\) and \(v_i(k)\) denote the clean speech and noise signals at channel \(i\), respectively. The clean speech is obtained by convolving the source signal \(s(k)\) with the \ac{HRIR} \(h_i(k)\), where \(*\) denotes the convolution. The corresponding STFT domain representation can be expressed as

{\ninept\begin{equation}
Y_i(t,f) = X_i(t,f) + V_i(t,f),
\label{eq:stft_domain_mixture}
\end{equation}}
where \(Y_i(t,f)\), \(X_i(t,f)\), and \(V_i(t,f)\) denote the STFT-domain signals at  time frame \(t\), and frequency bin \(f\) of the noisy mixture, clean speech, and noise signals, for channel \(i\), respectively. The DNN estimates a complex-valued mask \(\widehat{M}_i(t,f)\) for each channel, which is applied to \(Y_i(t,f)\) to obtain an estimate of clean speech \(X_i(t,f)\),
{\ninept\begin{equation}
\hat{X}_i(t,f) = \widehat{M}_i(t,f)\,Y_i(t,f).
\label{eq:masking}
\end{equation}}

\subsection{Baseline binaural cue loss}

The spatial information characterized by \ac{ILD} and \ac{IPD} in the T-F domain is given as
{\ninept\begin{equation}
\mathrm{ILD}_{X}(t,f)=20\log_{10}\left(\frac{|X_L(t,f)|}{|X_R(t,f)|}\right),
\label{eq:ild_def}
\end{equation}}
{\ninept\begin{equation}
\mathrm{IPD}_{X}(t,f) = \angle \left( \frac{X_L(t,f)}{X_R(t,f)} \right),
\label{eq:ipd_def}
\end{equation}}
where \(\angle(\cdot)\) denotes the phase operator.
Existing \ac{BSE} methods commonly assess spatial cue preservation using the weighted mean absolute errors \(\mathcal{L}_{\mathrm{ILD}}\) and \(\mathcal{L}_{\mathrm{IPD}}\) \cite{bc_transformer,bc_fourier}. \(\mathcal{L}_{\mathrm{ILD}}\) and \(\mathcal{L}_{\mathrm{IPD}}\) are given by
{\ninept\begin{equation}
\mathcal{L}_{\mathrm{ILD}} =
\frac{1}{N}
\sum_{t,f}
\mathcal{M}(t,f)
\left|
\mathrm{ILD}_{X}(t,f)-\mathrm{ILD}_{\hat{X}}(t,f)
\right|,
\label{eq:lild}
\end{equation}}
{\ninept\begin{equation}
\mathcal{L}_{\mathrm{IPD}} =
\frac{1}{N}
\sum_{t,f}
\mathcal{M}(t,f)
\left|
\mathrm{IPD}_{X}(t,f)-\mathrm{IPD}_{\hat{X}}(t,f)
\right|,
\label{eq:lipd}
\end{equation}}
where \(\mathcal{M}(t,f)\) is an \ac{IBM} as defined in \cite{bc_transformer}, selecting speech-active T-F bins so that errors are only computed in speech-active regions, and \(N=\sum_{t,f}\mathcal{M}(t,f)\). Prior work combines the \ac{ILD} and \ac{IPD} distortions into the cue loss \(\mathcal{L}_{\mathrm{ILD+IPD}}=\mathcal{L}_{\mathrm{ILD}}+10\mathcal{L}_{\mathrm{IPD}}\), and uses it as a training objective \cite{bc_transformer,bc_lstm,bc_fourier}. In this work, this combined cue loss is used as the baseline cue loss. From \cite{bc_transformer,bc_lstm,bc_fourier}, only the binaural cue terms are retained, since the additional \(\mathcal{L}_{\mathrm{SNR}}\) and \(\mathcal{L}_{\mathrm{STOI}}\) terms are not directly related to cue preservation. While \(\mathcal{L}_{\mathrm{ILD}}\) is effective in assessing \ac{ILD} preservation, \(\mathcal{L}_{\mathrm{IPD}}\) directly compares \ac{IPD} values, which can be unreliable due to phase wrapping effects. Two phases that are close on the unit circle but lie on opposite sides of the \(\pm \pi\) boundary may yield a large error under direct subtraction, even though they correspond to nearly identical spatial cue. 

\subsection{Proposed binaural cue loss}
\label{ssec:scp_loss}

We propose two binaural cue loss functions. The first is based on the \ac{RTF} and directly penalizes distortion in the interchannel relationship introduced by T-F masking. For this, we first define the time-invariant \ac{RTF}, which is obtained by least-squares estimation
{\ninept\begin{equation}
H_{\mathrm{LR}}(f)=\frac{\sum_t X_L(t,f)X_R^*(t,f)}{\sum_t |X_R(t,f)|^2},
\label{eq:hlr}
\end{equation}}
where \((\cdot)^*\) denotes complex conjugation. \(H_{\mathrm{RL}}(f)\) is defined analogously to \(H_{\mathrm{LR}}(f)\). In \ac{BSE}, T-F masking should suppress noise while preserving this interaural structure. If masking distorts this relation, the \ac{RTF} is altered. To measure this distortion, the estimated masks are first applied to the clean spectrum, with the time and frequency indices omitted where possible for brevity:
{\ninept\begin{equation}
\tilde{X}_L=\widehat{M}_L X_L, \qquad
\tilde{X}_R=\widehat{M}_R X_R.
\label{eq:masked_clean}
\end{equation}}
The masked \ac{RTF}s \(\tilde{H}_{\mathrm{LR}}(f)\) and \(\tilde{H}_{\mathrm{RL}}(f)\) are estimated in the same manner as \eqref{eq:hlr}. If the interaural structure is preserved after masking, one channel should still be reconstructable from the other using the corresponding masked \ac{RTF}s. Based on this principle, the binaural reconstruction error loss \(\mathcal{L}_{\mathrm{BRE}}\) is defined as
{\ninept\begin{equation}
\mathcal{L}_{\mathrm{BRE}}=
\mathrm{CCMSE}(X_L,\tilde{H}_{LR}X_R)
+
\mathrm{CCMSE}(X_R,\tilde{H}_{RL}X_L),
\label{eq:recon_metric}
\end{equation}}
where \(\mathrm{CCMSE}(\cdot,\cdot)\) denotes the complex compressed mean-squared error \cite{ccmse}. The \(\mathrm{CCMSE}\) is defined as
{\ninept\begin{equation}
\scalebox{0.9}{$
\mathrm{CCMSE}(\hat{X},X)
=
\left\|
|\hat{X}|^{\gamma} - |X|^{\gamma}
\right\|_2^2
+
\left\|
|\hat{X}|^{\gamma} e^{j\psi_{\hat{X}}}
-
|X|^{\gamma} e^{j\psi_X}
\right\|_2^2
$},
\label{eq:ccmse}
\end{equation}}
where \(\psi_{\hat X}\) and \(\psi_X\) are the phases of \(\hat X\) and \(X\), and \(\gamma = 0.3\) is a compression factor. The cue losses described in Section 2.2 is computed from the enhanced outputs and therefore does not isolate the distortion introduced by T-F masking itself. The motivation for using \(\mathcal{L}_{\mathrm{BRE}}\) is to decouple noise reduction from cue preservation by directly penalizing masking-induced distortion in the interchannel relationship, rather than relying on the enhanced output to preserve binaural cues.

As an alternate approach for preserving binaural cues, the second proposed loss jointly models \ac{ILD} and \ac{IPD} distortion. Motivated by the interchannel ratio, we first define a combined \ac{ILD}-\ac{IPD} representation as
{\ninept\begin{equation}
Q_X(t,f)=|\mathrm{ILD}_X(t,f)|e^{j\mathrm{IPD}_X(t,f)}.
\label{eq:q_def}
\end{equation}}
The proposed binaural cue loss is then given by
{\ninept\begin{equation}
\mathcal{L}_{\mathrm{BC}}
=
\frac{1}{N}
\sum_{t,f}
\mathcal{M}
\left(
\big||\mathrm{ILD}_{\hat X}|-|\mathrm{ILD}_X|\big|^2
+
\big|Q_{\hat X}-Q_X\big|^2
\right),
\label{eq:bc_loss}
\end{equation}}
where \(\mathcal{M}\) and \(N\) are defined in \eqref{eq:lild} and \eqref{eq:lipd}. The first term penalizes \ac{ILD} distortion, while the second term constrains the joint complex \ac{ILD}-\ac{IPD} representation. By encoding \ac{IPD} on the complex unit circle, the proposed loss avoids direct phase subtraction and thereby reduces sensitivity to phase wrapping.

\subsection{Noise reduction Loss}

The model is trained to optimize for noise reduction using multi-resolution spectrogram loss \cite{Deepfilternet}.
{\ninept\begin{equation}
\mathcal{L}_{\mathrm{NR}}=
\sum_i\sum_m
\mathrm{CCMSE}(\hat{X}'_{i,m},X'_{i,m}),
\label{eq:nr_loss}
\end{equation}}
where \(X'_{i,m}=\mathrm{STFT}_m(x_i)\), is the m-th STFT with window size in \(\{10,20,32,40\}\,\mathrm{ms}\), \(\hat{X}'_{i,m}\) is defined similarly for the enhanced signal \(\hat{x}_i\), and \(\mathrm{CCMSE}(\cdot,\cdot)\) as defined in \eqref{eq:ccmse}. The overall training objective is 
{\ninept\begin{equation}
\mathcal{L}=\mathcal{L}_{\mathrm{NR}}+\lambda\mathcal{L}_{\mathrm{spatial}},
\label{eq:total_loss}
\end{equation}}
where \(\mathcal{L}_{\mathrm{spatial}} \in \{\mathcal{L}_{\mathrm{BC}},\mathcal{L}_{\mathrm{BRE}}\}\), and \(\lambda\) is chosen so that both losses result in similar orders of magnitude. 


\section{Experiments}
In this section, we describe the DNN architecture, the data generation process for speech-in-noise samples, the training and evaluation setup, and the evaluation results of the proposed methods. 
\label{sec:experiments}

\subsection{Model architecture}
The DNN follows an encoder-GRU-decoder architecture as shown in Fig.~\ref{fig:block_diagram}. The encoder consists of six 2-D convolutional layers with output channels \(\{16,16,32,64,128,128\}\), kernel size \((5,3)\) and stride \((2,1)\).  The decoder mirrors the encoder using transposed Conv2D layers.  Two GRU layers with hidden size 128 are used in the bottleneck, followed by a linear layer. The input features are the \ac{STFT} signals of the left and right microphone signals, with magnitude compression \(c=0.3\) applied. The network estimates complex T-F masks for the left and right spectra.

\subsection{Data generation}
For training and validation, monaural speech signals are taken from the LibriSpeech \cite{Librispeech}, TIMIT \cite{timit_dataset}, and EARS \cite{ears_dataset} datasets which are spatialized using HRIRs from KEMAR head-and-torso \cite{HRIR_dataset}. The training and validation sets contain 90,000 and 10,000 utterances, respectively, each of \(\SI{5}{\second}\) duration. For testing, 2700 unseen utterances from LibriSpeech dataset \cite{Librispeech} is used. Noise signals are drawn from the NOISEX-92 database \cite{Noisex92} and used to synthesize isotropic diffuse noise fields by spatially distributing independent sources over the horizontal plane at azimuth intervals \(4^\circ\) using \ac{HRIR}s from \cite{HRIR_dataset}. For both training and evaluation, the target speech source is positioned in the front plane \(-90^\circ\) to \(+90^\circ\), to reflect realistic listening scenarios. The noise types used for training and evaluation are factory noise, pink noise, engine noise, speech-shaped noise, and white Gaussian noise. The time-domain signals are sampled at 16~kHz. The input signals are transformed to the STFT domain using a FFT size of 512 and a hop size of 256. The noise signals are added to speech at a target SNR from \(\{-5,0,5,10,15\}\)~dB.

\subsection{Training setup}
\label{sec:results}
\begin{figure*}[!t]
    \centering
    \includegraphics[width=\textwidth,height=25\textheight,keepaspectratio]{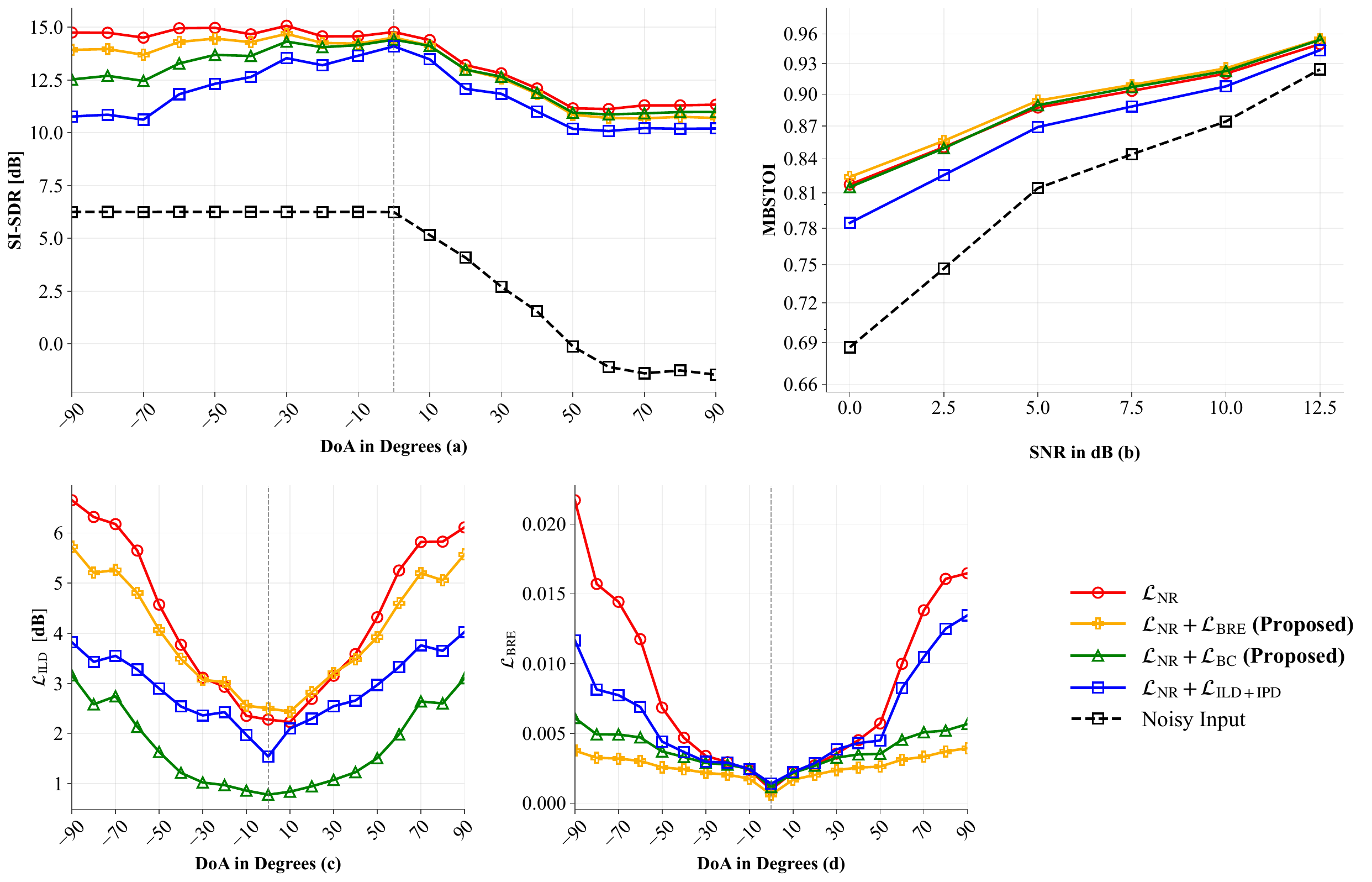}
    \caption{(a) Left-channel SI-SDR versus \ac{DoA}, averaged over SNRs; (b) MBSTOI versus SNR; and (c) \(\mathcal{L}_{\mathrm{ILD}}\) and (d) \(\mathcal{L}_{\mathrm{BRE}}\) versus \ac{DoA}, averaged over SNRs, for the different loss functions. For SI-SDR and MBSTOI, higher values indicate better performance, whereas for \(\mathcal{L}_{\mathrm{ILD}}\) and \(\mathcal{L}_{\mathrm{BRE}}\), lower values (\(\downarrow\)) indicate better performance.}
    \label{fig:cue_metrics}
\end{figure*}

 The network is trained for 60 epochs using the AdamW optimizer \cite{adam_w}. The initial learning rate is set to \(10^{-3}\), which is reduced by a factor of 0.1 on plateau, and the weight decay is set to \(10^{-5}\). For the IBM \(\mathcal{M}\), threshold is set to \(\Gamma=20\)~dB \cite{bc_transformer}. The proposed model contains approximately 0.8 million trainable parameters.  The cue loss weight \(\lambda\) is dynamically computed in each training step from the ratio of noise reduction loss and cue preservation loss term. 

\subsection{Evaluation setup}
Models trained with different spatial cue losses are evaluated in terms of both noise reduction and binaural cue preservation. For noise reduction, \ac{MBSTOI} \cite{mbstoi} and SI-SDR \cite{si_sdr} are reported. To assess binaural cue preservation, we report both \(\mathcal{L}_{\mathrm{ILD}}\) and \(\mathcal{L}_{\mathrm{BRE}}\). \(\mathcal{L}_{\mathrm{BRE}}\) is chosen as a cue evaluation metric as it provides an interpretable measure of masking-induced distortion in the interchannel relationship, incorporates phase information, and is less directly affected by overall denoising performance. The proposed cue losses \(\mathcal{L}_{\mathrm{BC}}\) and \(\mathcal{L}_{\mathrm{BRE}}\) are compared with a baseline trained using only \(\mathcal{L}_{\mathrm{NR}}\), and with a baseline trained using the baseline cue loss defined in Section~2.2, which serves as the state-of-the-art cue loss baseline in this work. All models are trained using the same noise reduction loss \(\mathcal{L}_{\mathrm{NR}}\) and differ only in the spatial loss term. During evaluation, the target SNR is selected uniformly from \(\{0, 2.5, 5, 7.5, 10, 12.5\}\,\mathrm{dB}\).
\subsection{Results and discussion}

Fig.~\ref{fig:cue_metrics} summarizes the results for the different evaluation metrics. Fig.~\ref{fig:cue_metrics}(a) shows the left-channel SI-SDR as a function of \ac{DoA}. The \(\mathcal{L}_{\mathrm{NR}}\)-only model achieves the best SI-SDR for all \ac{DoA}s, since it is optimized solely for denoising and is not constrained to preserve binaural cues. Both proposed cue losses, \(\mathcal{L}_{\mathrm{BC}}\) and \(\mathcal{L}_{\mathrm{BRE}}\), remain close to the \(\mathcal{L}_{\mathrm{NR}}\)-only baseline and outperform the baseline cue loss. The \(\mathcal{L}_{\mathrm{BRE}}\)-based model gives SI-SDR values close to the \(\mathcal{L}_{\mathrm{NR}}\)-only baseline because it penalizes only masking-induced distortion, without relying on the enhanced outputs to preserve binaural cues. The SI-SDR decreases as the source moves to the right, consistent with head shadowing. Fig.~\ref{fig:cue_metrics}(b) shows MBSTOI as a function of SNR. The same trend is observed, where both \(\mathcal{L}_{\mathrm{BC}}\)-based and \(\mathcal{L}_{\mathrm{BRE}}\)-based models have similar performance to the \(\mathcal{L}_{\mathrm{NR}}\)-only baseline and outperform the baseline cue loss.
Fig.~\ref{fig:cue_metrics}(c) shows \(\mathcal{L}_{\mathrm{ILD}}\) as a function of \ac{DoA}. The \ac{ILD} error increases toward the lateral directions (\(\pm 90^\circ\)), where level differences are larger. The \(\mathcal{L}_{\mathrm{NR}}\)-only model shows the worst cue preservation, highlighting the need for explicit cue-preserving objectives. The proposed \(\mathcal{L}_{\mathrm{BC}}\)-based model gives the lowest \ac{ILD} error and outperforms both the baseline cue loss and the \(\mathcal{L}_{\mathrm{NR}}\)-only model, indicating a better overall trade-off between noise reduction and cue preservation. In contrast, the \(\mathcal{L}_{\mathrm{BRE}}\)-based model shows poor \(\mathcal{L}_{\mathrm{ILD}}\) performance, since it does not use the enhanced outputs to preserve \ac{ILD}. Fig.~\ref{fig:cue_metrics}(d) shows \(\mathcal{L}_{\mathrm{BRE}}\) as a function of \ac{DoA}. As expected, the model trained with \(\mathcal{L}_{\mathrm{BRE}}\) achieves the lowest reconstruction error. The proposed \(\mathcal{L}_{\mathrm{BC}}\)-based model also achieves lower reconstruction error than the baseline cue loss and the \(\mathcal{L}_{\mathrm{NR}}\)-only model, indicating better preservation of the interchannel relationship. Together, these results suggest that no single metric fully captures binaural cue preservation, since the two metrics emphasize different aspects of the binaural structure.

\section{Conclusion}
\label{conclusion}
In this work, we introduced two binaural cue loss functions. The first, \(\mathcal{L}_{\mathrm{BRE}}\), directly penalizes masking-induced distortion in the interchannel relationship. The second, \(\mathcal{L}_{\mathrm{BC}}\), jointly constrains \ac{ILD} and \ac{IPD} in a unified representation. Experiments conducted under controlled anechoic conditions showed that the proposed \(\mathcal{L}_{\mathrm{BC}}\) achieves lower interaural level difference error than the baseline cue loss while maintaining competitive noise reduction performance. The proposed losses also reduce binaural reconstruction error, indicating less masking-induced distortion of the binaural structure. Future work will extend this study to reverberant environments for a more robust analysis.

\begingroup
\small
\bibliographystyle{IEEEbib}
\bibliography{refs}

@INPROCEEDINGS{Deepfilternet,
  author={Schröter, H. and Maier, A. and Escalante-B, A.N. and Rosenkranz, T.},
  booktitle={2022 International Workshop on Acoustic Signal Enhancement (IWAENC)}, 
  title={Deepfilternet2: Towards Real-Time Speech Enhancement on Embedded Devices for Full-Band Audio}, 
  year={2022},
  volume={},
  number={},
  pages={1-5},
  doi={10.1109/IWAENC53105.2022.9914782}}

@inproceedings{dccrn,
  title={DCCRN: Deep Complex Convolution Recurrent Network for Phase-Aware Speech Enhancement},
  author={Yanxin Hu and Yun Liu and Shubo Lv and Mengtao Xing and Shimin Zhang and Yihui Fu and Jian Wu and Bihong Zhang and Lei Xie},
  booktitle={Interspeech},
  year={2020},
  url={https://api.semanticscholar.org/CorpusID:220936516}
}

@article{HRIR_dataset,
  title = {A Multiple Model High-Resolution Head-Related Impulse Response Database for Aided and Unaided Ears},
  author = {Thiemann, Joachim and Van De Par, Steven},
  year = 2019,
  month = dec,
  journal = {EURASIP Journal on Advances in Signal Processing},
  volume = {2019},
  number = {1},
  pages = {9},
  issn = {1687-6180},
  doi = {10.1186/s13634-019-0604-x},
  urldate = {2026-04-23},
  langid = {english}
}

@inproceedings{Librispeech,
  title = {Librispeech: {{An ASR}} Corpus Based on Public Domain Audio Books},
  shorttitle = {Librispeech},
  booktitle = {2015 {{IEEE International Conference}} on {{Acoustics}}, {{Speech}} and {{Signal Processing}} ({{ICASSP}})},
  author = {Panayotov, Vassil and Chen, Guoguo and Povey, Daniel and Khudanpur, Sanjeev},
  date = {2015-04},
  pages = {5206--5210},
  publisher = {IEEE},
  location = {South Brisbane, Queensland, Australia},
  doi = {10.1109/ICASSP.2015.7178964},
  url = {http://ieeexplore.ieee.org/document/7178964/},
  urldate = {2026-03-20},
  eventtitle = {{{ICASSP}} 2015 - 2015 {{IEEE International Conference}} on {{Acoustics}}, {{Speech}} and {{Signal Processing}} ({{ICASSP}})},
  isbn = {978-1-4673-6997-8}
}

@article{timit_dataset,
author = {Garofolo, J. and Lamel, Lori and Fisher, W. and Fiscus, Jonathan and Pallett, D. and Dahlgren, N. and Zue, V.},
year = {1992},
month = {11},
pages = {},
title = {TIMIT Acoustic-phonetic Continuous Speech Corpus},
journal = {Linguistic Data Consortium}
}

@inproceedings{ears_dataset,
  title={{EARS}: An Anechoic Fullband Speech Dataset Benchmarked for Speech Enhancement and Dereverberation},
  author={Richter, Julius and Wu, Yi-Chiao and Krenn, Steven and Welker, Simon and Lay, Bunlong and Watanabe, Shinjii and Richard, Alexander and Gerkmann, Timo},
  booktitle={Interspeech},
  year={2024}
}

@article{nr_ha1,
  title={Noise reduction in hearing aids: a review.},
  author={Levitt, Harry},
  journal={Journal of Rehabilitation Research \& Development},
  volume={38},
  number={1},
  year={2001}
}

@article{nr_ha2,
author = {Levitt, Harry and Bakke, M and Kates, James and Neuman, Arlene and Schwander, T and Weiss, M},
year = {1993},
month = {02},
pages = {7-19},
title = {Signal processing for hearing impairment},
volume = {38},
journal = {Scandinavian audiology. Supplementum}
}

@INPROCEEDINGS{nr_ar,
  author={Guiraud, Pierre and Hafezi, Sina and Naylor, Patrick A. and Moore, Alastair H. and Donley, Jacob and Tourbabin, Vladimir and Lunner, Thomas},
  booktitle={2022 International Workshop on Acoustic Signal Enhancement (IWAENC)}, 
  title={An Introduction to the Speech Enhancement for Augmented Reality (Spear) Challenge}, 
  year={2022},
  volume={},
  number={},
  pages={1-5},
  doi={10.1109/IWAENC53105.2022.9914721}}

@inbook{beamforming,
author = {Doclo, Simon and Gannot, Sharon and Moonen, Marc and Spriet, Ann},
publisher = {John Wiley \& Sons, Ltd},
isbn = {9780470487068},
title = {Acoustic Beamforming for Hearing Aid Applications},
booktitle = {Handbook on Array Processing and Sensor Networks},
chapter = {9},
pages = {269-302},
doi = {https://doi.org/10.1002/9780470487068.ch9},
url = {https://onlinelibrary.wiley.com/doi/abs/10.1002/9780470487068.ch9},
eprint = {https://onlinelibrary.wiley.com/doi/pdf/10.1002/9780470487068.ch9},
year = {2010}
}

@ARTICLE{spectral_substraction,
  author={Boll, S.},
  journal={IEEE Transactions on Acoustics, Speech, and Signal Processing}, 
  title={Suppression of acoustic noise in speech using spectral subtraction}, 
  year={1979},
  volume={27},
  number={2},
  pages={113-120},
  doi={10.1109/TASSP.1979.1163209}}

@INPROCEEDINGS{bc_transformer,
  author={Tokala, Vikas and Grinstein, Eric and Brookes, Mike and Doclo, Simon and Jensen, Jesper and Naylor, Patrick A.},
  booktitle={ICASSP 2024 - 2024 IEEE International Conference on Acoustics, Speech and Signal Processing (ICASSP)}, 
  title={Binaural Speech Enhancement Using Deep Complex Convolutional Transformer Networks}, 
  year={2024},
  volume={},
  number={},
  pages={681-685},
  doi={10.1109/ICASSP48485.2024.10447090}}

@article{Noisex92,
title = {Assessment for automatic speech recognition: II. NOISEX-92: A database and an experiment to study the effect of additive noise on speech recognition systems},
journal = {Speech Communication},
volume = {12},
number = {3},
pages = {247-251},
year = {1993},
issn = {0167-6393},
doi = {https://doi.org/10.1016/0167-6393(93)90095-3},
url = {https://www.sciencedirect.com/science/article/pii/0167639393900953},
author = {Andrew Varga and Herman J.M. Steeneken},
}

@article{td_1,
   title={Conv-TasNet: Surpassing Ideal Time–Frequency Magnitude Masking for Speech Separation},
   volume={27},
   ISSN={2329-9304},
   url={http://dx.doi.org/10.1109/TASLP.2019.2915167},
   DOI={10.1109/taslp.2019.2915167},
   number={8},
   journal={IEEE/ACM Transactions on Audio, Speech, and Language Processing},
   publisher={Institute of Electrical and Electronics Engineers (IEEE)},
   author={Luo, Yi and Mesgarani, Nima},
   year={2019},
   month=aug, pages={1256–1266} }

@inproceedings{td_2,
  author       = {Daniel Stoller and
                  Sebastian Ewert and
                  Simon Dixon},
  title        = {Wave-U-Net: {A} Multi-Scale Neural Network for End-to-End Audio Source
                  Separation},
  booktitle    = {Proceedings of the 19th International Society for Music Information
                  Retrieval Conference, {ISMIR} 2018, Paris, France, September 23-27,
                  2018},
  pages        = {334--340},
  year         = {2018},
  url          = {http://ismir2018.ircam.fr/doc/pdfs/205\_Paper.pdf},
  bibsource    = {dblp computer science bibliography, https://dblp.org}
}

@INPROCEEDINGS{bc_fourier,
  author={Lu, Xikun and Ma, Yujian and Jiang, Xianquan and Wang, Xuelong and Sang, Jinqiu},
  booktitle={ICASSP 2026 - 2026 IEEE International Conference on Acoustics, Speech and Signal Processing (ICASSP)}, 
  title={A Lightweight Fourier-Based Network for Binaural Speech Enhancement with Spatial Cue Preservation}, 
  year={2026},
  volume={},
  number={},
  pages={12227-12231},
  doi={10.1109/ICASSP55912.2026.11463903}}

@INPROCEEDINGS{bc_old,
  author={Tokala, Vikas and Brookes, Mike and Naylor, Patrick A.},
  booktitle={2022 International Workshop on Acoustic Signal Enhancement (IWAENC)}, 
  title={Binaural Speech Enhancement Using STOI-optimal Masks}, 
  year={2022},
  volume={},
  number={},
  pages={1-5},
  doi={10.1109/IWAENC53105.2022.9914744}}

@inproceedings{bc_lstm,
   title={Binaural Speech Enhancement Using Complex Convolutional Recurrent Networks},
   url={http://dx.doi.org/10.1109/IEEECONF59524.2023.10476738},
   DOI={10.1109/ieeeconf59524.2023.10476738},
   booktitle={2023 57th Asilomar Conference on Signals, Systems, and Computers},
   publisher={IEEE},
   author={Tokala, Vikas and Grinstein, Eric and Brookes, Mike and Doclo, Simon and Jensen, Jesper and Naylor, Patrick A.},
   year={2023},
   month=oct, pages={1130–1134} }

@INPROCEEDINGS{si_sdr,
  author={Roux, Jonathan Le and Wisdom, Scott and Erdogan, Hakan and Hershey, John R.},
  booktitle={ICASSP 2019 - 2019 IEEE International Conference on Acoustics, Speech and Signal Processing (ICASSP)}, 
  title={SDR – Half-baked or Well Done?}, 
  year={2019},
  volume={},
  number={},
  pages={626-630},
  doi={10.1109/ICASSP.2019.8683855}}

@INPROCEEDINGS{ccmse,
  author={Braun, Sebastian and Gamper, Hannes and Reddy, Chandan K.A. and Tashev, Ivan},
  booktitle={ICASSP 2021 - 2021 IEEE International Conference on Acoustics, Speech and Signal Processing (ICASSP)}, 
  title={Towards Efficient Models for Real-Time Deep Noise Suppression}, 
  year={2021},
  volume={},
  number={},
  pages={656-660},
  doi={10.1109/ICASSP39728.2021.9413580}}

@article{mbstoi,
title = {Refinement and validation of the binaural short time objective intelligibility measure for spatially diverse conditions},
journal = {Speech Communication},
volume = {102},
pages = {1-13},
year = {2018},
issn = {0167-6393},
doi = {https://doi.org/10.1016/j.specom.2018.06.001},
url = {https://www.sciencedirect.com/science/article/pii/S0167639317302947},
author = {Asger Heidemann Andersen and Jan Mark {de Haan} and Zheng-Hua Tan and Jesper Jensen}
}

@inproceedings{adam_w,
  title={Decoupled Weight Decay Regularization},
  author={Ilya Loshchilov and Frank Hutter},
  booktitle={International Conference on Learning Representations},
  year={2017},
  url={https://api.semanticscholar.org/CorpusID:53592270}
}
\endgroup

\end{document}